\documentclass[superscriptaddress,showkeys]{revtex4-2}

\usepackage{graphics,epsfig }
 
 \usepackage{xcolor}
\usepackage{footmisc}
\usepackage{graphicx}
\usepackage{dcolumn}
\usepackage{amsmath}
\usepackage{hyperref}
\newcommand{\be}{\begin{equation}}
\newcommand{\ee}{\end{equation}}
\newcommand{\bea}{\begin{eqnarray}}
\newcommand{\eea}{\end{eqnarray}}

\def\be{\begin{equation}}
\def\ee{\end{equation}}
\def\ba{\begin{eqnarray}}
\def\ea{\end{eqnarray}}

\begin{document}
\title{Quasinormal modes, shadow and thermodynamics of  black holes coupled with nonlinear electrodynamics and cloud of strings }

\author{Dharm Veer Singh\footnote{Visiting Associate, IUCAA  
Pune, Maharashtra 411007, India\label{note3}}}
\email{veerdsingh@gmail.com; dharm.singh@associates.iucaa.in}
\affiliation{Department Physics,
 GLA University, Mathura 281406 Uttar Pradesh}

\author{Aradhya Shukla}
\email{ashukla038@gmail.com}
\affiliation{Department Physics,
 GLA University, Mathura 281406 Uttar Pradesh}

\author{Sudhaker Upadhyay\footnote{Corresponding author}\footref{note3}  
}
\email{sudhakerupadhyay@gmail.com; sudhaker@associates.iucaa.in}
\affiliation{Department of Physics, K.L.S. College,  Magadh University, Nawada, Bihar 805110, India}

\affiliation{School of Physics, Damghan University, P.O. Box 3671641167, \\Damghan,  Iran}

\begin{abstract}
\noindent We construct an exact  black hole solution for the Einstein gravity coupled with the nonlinear electrodynamics (which corresponds to the Maxwell electrodynamics in the weak field limit) in the presence of a cloud of strings as the source.  We study the thermodynamical properties of the black hole solutions and derive the corrected 
first-law of thermodynamics. The presence of a cloud of strings does not affect the stability of the present black hole. However, a second-order phase transition exists for this system at a critical horizon radius. Furthermore, we study
the   quasinormal modes and their shadow radius. In addition, we find that, upon variation,  the parameters
of the theory show different aspects of the optical characteristics of the black hole solutions.
\end{abstract} 
\maketitle

\section{Introduction}\label{sec:level1}
One of the important aspects of general relativity is to find solutions to Einstein's field equation.  The
 metric tensor in arbitrarily chosen spacetimes may lead to an unphysical
stress-tensor. Therefore, obtaining a meaningful 
 solution to Einstein's field equation is a cumbersome job \cite{st}. 
The well-known vacuum (without matter) solution and with matter solution of Einstein's field
equation is the Schwarzschild black hole and Reissner-Nordstr\"om black hole, respectively. These spherically symmetric solutions have a central singularity which is separated by a boundary known as the event horizon.   
 On the other hand, non-singular solutions of  Einstein's  field equations   are known as regular black holes. Indeed, one can have   black hole metrics without a physical singularity. Bardeen was the first to realize   the idea of a
central matter core  and to propose a regular black hole  solution   with horizons (but without a singularity) \cite{bar}. The Bardeen regular black hole is a spherically symmetric solution that violates the strong energy condition. Furthermore, other spherically symmetric regular solutions   were proposed  \cite{reg,reg1,reg2}.  
There has been
tremendous development in the investigations of  regular black hole solutions and their properties \cite{gla,gla1,gla2,gla3}.
 Recently, it is observed  that regular black hole solutions    with spherical symmetry violet
the weak energy condition \cite{na}. 

It has been observed that   the
linearity of the real electromagnetic field  breaks at high energies due to interactions with other physical fields. In such circumstances, the most suitable alternative  is  nonlinear electrodynamics
  as a simplified phenomenological description of this interaction. 
  The properties of the nonlinear electrodynamics  can be lucid from the perspective of gravity since strong vector fields are
dominant   at the center of black holes.
The behavior of charged particles around black holes  may be described by 
nonlinear electrodynamics. Hence, it is crucial to emphasize the
gravitational field coupled to nonlinear electrodynamics.  Born and Infeld 
were the first to propose nonlinear electrodynamics with the motivation to have
 a finite self-energy of the point charge \cite{born}.  Moreover, a lot of 
 progress have been made in various contexts \cite{b1,b2,b3,b4,b5}.    Nonlinear electrodynamics also gets relevance as
  a generalized Born-Infeld action appears
  naturally   from the  effective string action \cite{b6,b7}.

  Letelier proposed a model analogous to pressure-less perfect fluid known as  cloud of strings  (CS) \cite{lat}. From this perspective,  
 a generalized Schwarzschild black hole  solution is obtained  in the background of a spherically symmetric cloud of strings   \cite{lat1,lat2}.   
  However,  the energy-momentum tensor  of the string  manifests a non-null pressure, which therefore leads to  astrophysical and
cosmological implications.  Further studies of a cloud of strings  in the general relativity and alternative theories of
gravity are reported in Refs. \cite{cs1,cs2,cs3,cs4,cs5,cs6}. Since strings   are supposed to be  fundamental constituents of the universe  rather than point-like particles.   This derives us to study the black hole solutions in presence of CS. More precisely, we provide an exact solution for
the Einstein gravity coupled to nonlinear electrodynamics  
in presence of a cloud of strings. We also discuss its thermodynamics, stability
and phase transition. 

Let us briefly introduce CS as matter source. The CS is governed by the Nambu-Goto action, which takes the form 
\begin{equation}  
 S_{\text{NG}} =\int_{\Sigma}  \; m (-\gamma)^{-1/2}  d\lambda^{0} d\lambda^{1}= \int_{\Sigma}  \; m \left(-\frac{1}{2} \Sigma^{\mu \nu} \Sigma_{\mu \nu}\right)^{1/2}  d\lambda^{0} d\lambda^{1},
\end{equation}
where constant $m$  characterizes each
string, ($\lambda^{0},  \lambda^{1}$) are  local coordinates of the string being timelike and  spacelike in nature, respectively \cite{synge}.   Here,   $\gamma $ refers to the determinant of an induced metric $\gamma_{a b}:=  g_{\mu \nu} \frac{\partial x^{\mu}}{\partial \lambda^{a}} \frac{\partial x^{\nu}}{\partial \lambda^{b}}$ on the strings world sheet.
  The bivector related to the string world sheet $\Sigma$ is  written by
\begin{equation}
\label{eq:bivector}
\Sigma^{\mu \nu} = \epsilon^{a b} \frac{\partial x^{\mu}}{\partial \lambda^{a}} \frac{\partial x^{\nu}}{\partial \lambda^{b}},
\end{equation}
where second rank Levi-Civita tensor $\epsilon^{a b}$ (anti-symmetric in $a$ and $b$) takes following non-zero values: $\epsilon^{0 1} = - \epsilon^{1 0} = 1$.

Using definition, $T^{\mu \nu} = 2 \frac{\partial }{\partial g_{\mu \nu}} m \left(-\frac{1}{2} \Sigma^{\mu \nu} \Sigma_{\mu \nu}\right)^{1/2}$, we obtain single string energy-momentum tensor     as $T^{\mu \nu} = \frac{m \Sigma^{\mu \rho} \Sigma_{\rho}^{\phantom{\rho} \nu}}{\sqrt{-\gamma}}$. Since  the CS
is   characterized by a proper density $\rho$ rather than a single mass, therefore,  
the energy-momentum tensor for CS is given by 
\begin{equation}
T^{\mu \nu}  = \frac{\rho \Sigma^{\mu \rho} \Sigma_{\rho}^{\phantom{\rho} \nu}}{\sqrt{-\gamma}}.
\end{equation}
Here,  $\rho \; (\gamma)^{-1/2} $ denotes a gauge-invariant density. 

The only surviving component of the bivector $\Sigma$ regarding spherically symmetric solution is $\Sigma^{tr} =  - \Sigma^{rt}$.  Thus,  conservation law leads to  
\begin{equation}
T^t_t = T^r_r = \frac{a}{r^2},
\label{emt2}
\end{equation}
for some real constant $a$, which is related to the global monopole charge\cite{mb}.

The  proper solutions of the perturbation equations of a black hole
belonging to certain complex characteristic frequencies (which satisfy  the boundary conditions) are described by quasinormal modes (QNMs)  \cite{chan}.
A close connection exists between QNMs and the shadow radius  of a black hole \cite{chan01}. The QNMs are determined entirely by the dynamics of black holes. At the same time, the shadow radius belongs to optical properties, such a connection  justifies  the relation between dynamics and  optical properties of black holes \cite{76,Jafarzade:2020ova}.
Recently, the shadow of the $5D$ AdS Reissner-Nordstr\"om  black hole is studied in Ref. \cite{sudhak}.

The paper is structured as follows. In section \ref{sec2}, we consider an action describing Einstein's gravity with nonlinear electrodynamics in the presence of  CS. We obtain  an exact  black hole solution with coordinate singularity for this model in section \ref{sec3}. Here, we discuss  the energy conditions for this black hole solution. {  Thermodynamical behavior of this black hole is presented in section \ref{sec03}. We see that the black hole  follows a modified first-law of thermodynamics. We study the photon radius including shadow radius in section \ref{sec04}. Within the section, we consider a photon  moving in a circular orbit. The photon radius  is obtained numerically.  The section \ref{sec05} is devoted to the study of the connection between shadow radius and QNMs.  We conclude the results and their importance in the last section. }

\section{Gravity coupled with nonlinear electrodynamics and cloud of strings}\label{sec2}
Now, the action describing Einstein's gravity coupled to nonlinear electrodynamics and  surrounded by a CS in $4$ dimensions is given by 
\begin{equation}
S=\frac{1}{2}\int_{\mathcal{M}}d^{4}x\sqrt{-g}\left[  R  +\mathcal{L}( {\mathcal F}) \right]+S_{\text{NG}},
\end{equation}
the Lagrangian $\mathcal{L}({\mathcal F})$ describes nonlinear electrodynamics   with invariant $ {\mathcal F}=\frac{1}{4}F_{\mu\nu}F^{\mu\nu}$,  where $F_{\mu\nu}= \nabla_ \mu A_{\nu}-\nabla_ \nu A_{\mu}$ is the electromagnetic field strength tensor for the gauge potential $A_{\nu}$.   

 The equations of motion  for the action are given by  \cite{sus}%
\begin{eqnarray}\label{ee}
&&G_{\mu\nu}\equiv R_{\mu\nu}-\frac{1}{2} g_{\mu\nu} R = {\mathcal T_{\mu\nu}} +T_{\mu\nu},\\
&&\nabla_{\mu}\left(\frac{\partial \mathcal{L(F)}}{\partial  {\mathcal F}}F_{\mu\nu}\right)=0~~~ \text{and}~~~\nabla_{\mu}\left(^*F_{\mu\nu}\right)=0.
\label{eom11}
\end{eqnarray}
where $G_{\mu\nu}$ is the Einstein tensor and  ${\mathcal T_{\mu\nu}} \equiv 2\left[\frac{\partial \mathcal{L} }{\partial  {\mathcal F}}F_{\mu\rho}F_{\nu}^{\rho}-g_{\mu\nu}\mathcal{L} \right]$ is energy-momentum tensor   related to the electromagnetic tensor.
 The explicit form of  Lagrangian density of the nonlinear electrodynamics is considered as \cite{Ghosh:2018bxg}
\begin{eqnarray}
\mathcal{L}( {\mathcal F}) ={\mathcal F} e^{-\frac{k}{q}(2q^2F)^{\frac{1}{4}}}, \qquad \text{with}\qquad k=\frac{q^2 }{2M}.\label{44}
\end{eqnarray}
Here, $q$ corresponds to the nonlinear charge of a self-gravitating magnetic field and $M$ is the mass parameter, later which is related to the mass of the black hole.  
The Lagrangian $\mathcal{L}({\mathcal F})$  is the function of ${\mathcal F}$, where
the Lagrangian $\partial \mathcal{L} /\partial {\mathcal F}\to \infty$ as ${\mathcal F}\to \infty$ and $\partial \mathcal{L}  /\partial {\mathcal F}\to 1$ as ${\mathcal F}\to 0$. In weak field limit (${\mathcal F} <<1 $), the $\mathcal{L}$ identifies to the  Maxwell electrodynamics. Nonetheless, under the strong field limit, $\mathcal{L}$ vanishes. 

We consider the following \textit{anstaz} for the Maxwell field \cite{ak19,sabir}:
\begin{eqnarray}
F_{\mu\nu}&=&2\delta^{\theta_{1}}_{[\mu}\delta^{\theta_{2}}_{\nu]}q\sin\theta.
\label{ee3}
\end{eqnarray}
with $ {\mathcal F}$ as
\begin{equation}
 {\mathcal F}=\frac{q^{2 }}{2r^{4}}.
\end{equation}
  The energy-momentum tensor, in this case, is  calculated by
\begin{eqnarray}\label{em}
{\mathcal T}^t_t &=& {\mathcal T}^r_r =\frac{2Mke^{-\frac{q^2}{2M r }}}{r^{4}},
\label{emt}
\end{eqnarray}
which satisfies the equations of motion in the case of nonlinear electrodynamics. 
 

\section{Regular black hole solutions with cloud of strings}\label{sec3}
\begin{figure*}[ht]
\begin{tabular}{c c c c}
\includegraphics[width=0.5\linewidth]{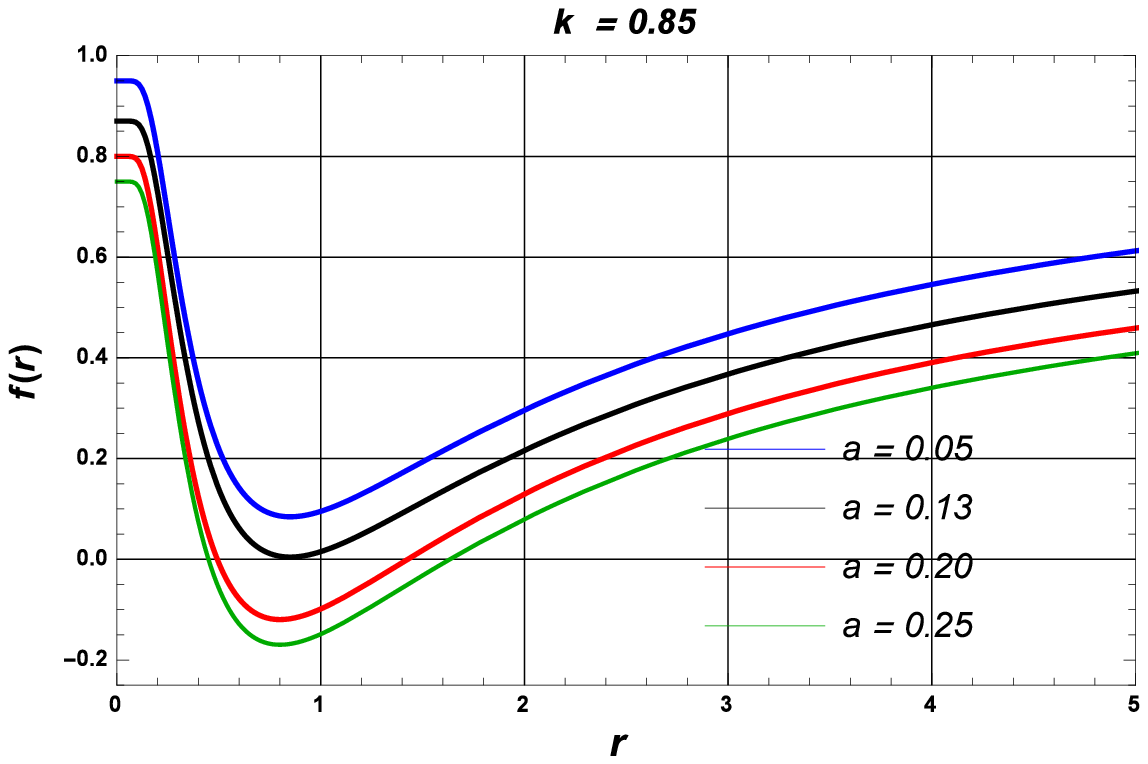}
\includegraphics[width=0.5\linewidth]{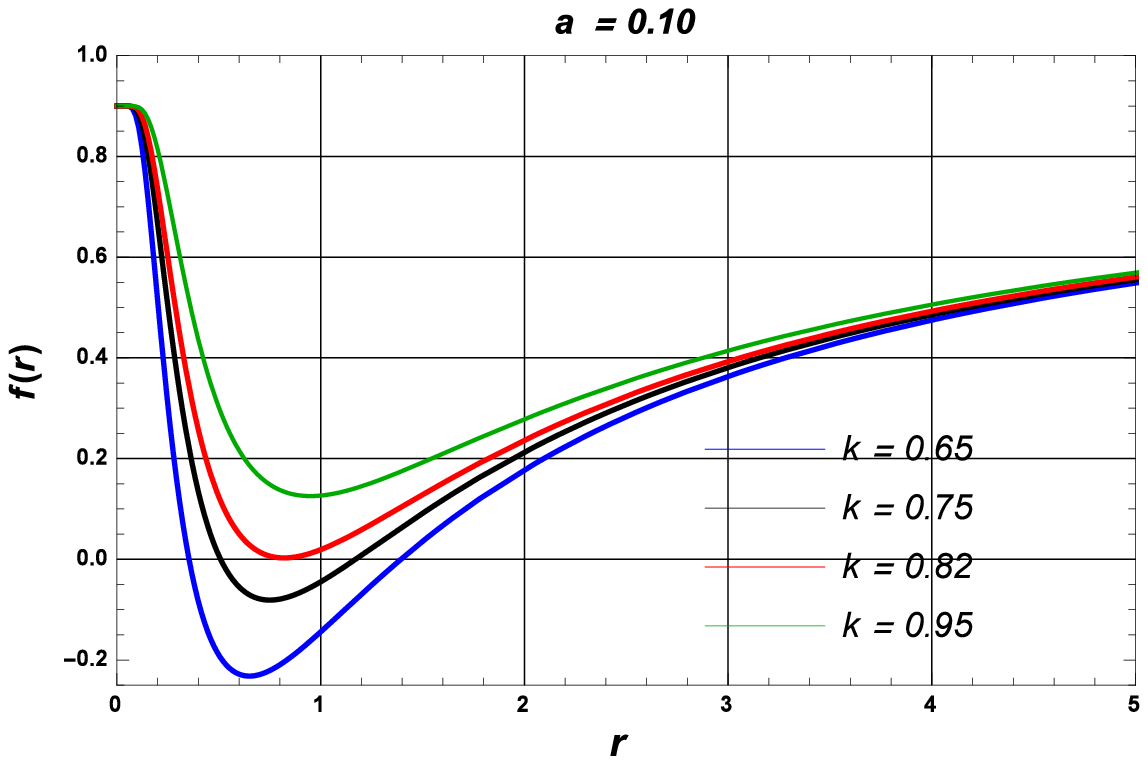}
\end{tabular}
\caption{The plot of metric function $f(r)$ versus horizon radius $r$ for different value of CS parameter $a$ (in the left plot) and  deviation parameter  $k$ (in the right plot).}
\label{m1}
\end{figure*}

In this section, the main motivation  is to get a static spherically symmetric solution  of the equation of motion (\ref{ee})  with a CS  and nonlinear electrodynamics as source  and investigate its properties. In order to achieve the goal, let us begin by writing spherically symmetric static metric of the form:
\begin{equation}
ds^2 = -f(r) dt^2+ \frac{1}{f(r)} dr^2 + r^2 d\Omega^2,
\label{met1}
\end{equation}
with
\begin{equation}
f(r)=1-\frac{{ 2}m(r)}{r},
\end{equation}
where $d\Omega^2=d\theta^2+\sin^2\theta d\phi^2$ is the metric on $2D$ sphere. Using   Eq. (6) with the metric element (13)
leads to the Einstein field equations to have the following form:
 \begin{eqnarray}
\frac{d}{dr}m(r)  =\frac{q^2}{2r^2}e^{-k/r}+\frac{a}{2}.
\label{eom1}
\end{eqnarray}
  Integrating the above expression  (\ref{eom1}), we have
  the explicit form of single string mass
\begin{equation}
m(r)=Me^{-k/ r}+\frac{a}{2}r +C_1,
\end{equation}
where $C_1$ is constant, then  $C_1=\lim_{r\to \infty} m(r)-ar=M$  and substituting the $m(r)$ into $f(r)$, the black hole solution (\ref{met1}) becomes
\begin{equation}
ds^2=-\left[1-\frac{2M}{r}e^{-k/r}-a\right]dt^2+\frac{1}{\left[1-\frac{2M}{r}e^{-k/r}-a\right]}dr^2+r^2d\Omega^2.
\label{sol1}
\end{equation}
 This is an exact black hole solution in the presence of  nonlinear source  $e^{-k/r}$  and CS parameter $a$. This black hole is characterized by the parameters like   $M$,   $k$ and  $a$. The presence of such a nonlinear source    and CS parameter ensure deviation from  Schwarzschild black hole.  In  the 
 limit $k=0$, the resulting solution  reduces to Letelier solution \cite{lat2}. However, this coincides  to Schwarzschild 
 black hole solution for $k=0$ and $a=0$. The solution  behaves as the  Reissner-Nordstr\"om black hole for the case of   $r>>k$. {  For  $a=0$ and $k<0$,   $m(r)$ increases exponentially for small $r$. On the other hand, when $k>0$,  $m(r)$ has the following properties}:
\begin{equation}
\lim_{r\to 0^{+}}e^{-k/r}+a=a, \qquad\qquad \lim_{r\to 0^{-}}e^{-k/r}+a=+ \infty.
\end{equation}
The function is discontinuous at $r=0$. For $r>>k$, the obtained solution   behaves as a Reissner-Nordstr\"om black hole with CS parameter $a$
as
\begin{equation}
f(r)=1-\frac{2M}{r}+\frac{q^2}{r^2}-a+{\cal O}\left(\frac{k^2}{r^2}\right).
\label{sol2}
\end{equation}
The charge $q$  and mass $M$ are related by the relation $q^2=2Mk$ as mentioned in (\ref{44}). It is not cumbersome to estimate the numerical range of  $M$,  $a$ and  $k$ for the  black hole solution  (\ref{sol1}). This  black hole has  two horizons, the  event horizon ($r_+$) and the Cauchy horizon ($r_-$).  These  horizons in terms of the Lambert $W$ function are given by
\begin{equation}
r_{\pm}=-\frac{k}{\text{W}\left(-\frac{(1-a)k}{{2}M}\right)}\qquad\implies \qquad r_{\pm}= \frac{{2}M}{(1-a)}e^{W\left(-\frac{(1-a)k}{M}\right)}.
\end{equation}
  The Lambert $W$ function has two branches $W_0$ and $W_{-1}$ and provides two possibilities,
\begin{eqnarray}
&&W_0\left(-\frac{(1-a)k}{{2}M}\right)<0\qquad \implies \qquad k\in \left(0,\frac{{2}M}{(1-a)q}\right],\nonumber\\
&&W_{-1}\left(-\frac{(1-a)k}{{2}M}\right)<0,\,\,\quad \implies \qquad k\in \left[-\frac{{2}M}{(1-a)q,0}\right).
\end{eqnarray}
The value of $k$ lies in this interval as we have well defined coordinate location for a horizon when taking the $\left(-{(1-a)k}/{M}\right)$ branch of the Lambert function. The range $W_{-1}$  branch is entirely $-ve$  and  only return output occurs when $\left[-{M}/{(1-a)q,0}\right)$. This means that   all the possible solutions will correspond  to $r_+>0$. The inner and event horizons are located at, respectively,
\begin{equation}
r_-=\frac{{2}M}{(1-a)}e^{W_{-1}\left(-\frac{(1-a)k}{{2}M}\right)},\qquad \text{and}\qquad r_+=\frac{{2}M}{(1-a)}e^{W_0\left(-\frac{(1-a)k}{{2}M}\right)}.
\end{equation}
Only for $k={2}M/(1-a)q$, the inner and event horizon coincides and we have an extremal black hole. Moreover, for  $k>{2}M/(1-a)q$, the Lambert $W$ function is undefined. In order to study the behavior of $f(r)$ as a function of the different parameters, we plot figure \ref{m1}.
However, the numerical analysis of the horizon is tabulated in
TABLE \ref{tab1}.
\begin{center}
\begin{table}[h]
\begin{center}
\begin{tabular}{l|l r l| r l r l r}
\hline
\hline
\multicolumn{1}{c|}{ }&\multicolumn{1}{c}{ }&\multicolumn{1}{c}{$k=0.85$  }&\multicolumn{1}{c|}{ \,\,\,\,\,\, }&\multicolumn{1}{c}{ }&\multicolumn{1}{c}{}&\multicolumn{1}{c}{ $a=0.10$ }&\multicolumn{1}{c}{}\,\,\,\,\,\,\\
\hline
\multicolumn{1}{c|}{ \it{a}} & \multicolumn{1}{c}{ $r_-$ } & \multicolumn{1}{c}{ $r_+$ }& \multicolumn{1}{c|}{$\delta$}&\multicolumn{1}{c}{\it{ k}}& \multicolumn{1}{c}{$r_-$} &\multicolumn{1}{c}{$r_+$} &\multicolumn{1}{c}{$\delta$}   \\
\hline

\,\,\,\, 0.13\,\, &  \,\,0.875\,\,  &\,\,0.875\,\,&\,\,0\,\,&0.65&\,\,0.357\,\,&\,\,1.39\,\,&\,\,1.037\,\,
 \\
\
\,\, 0.20\,\,& \,\,0.587\,\, &\,\,  1.313\,\,& \,\,0.726\,\,&0.75&\,\, 0.518\,\,&\,\,1.175\,\,&\,\,0.657\,\,
\\
\
\,\, 0.25\,\, & \,\,0.518\,\, &\,\,  1.509\,\,& \,\,0.991\,\,&0.813&\,\,0.904\,\,&\,\,0.904\,\,&\,\,0\,\,
\\
\hline
 \hline
\end{tabular}
\end{center}
\caption{Radius of inner and outer horizons and $\delta=r_+-r_-$ for different values of charge $q$.}
\label{tab1}
\end{table}
\end{center}
From the figure \ref{m1}, we can see that the metric function   of the black hole increases with increasing CS parameter $(a)$  and the metric of the black hole decreases with  increasing   deviation parameter ($k$). It is worthwhile to mention that the behavior of the CS parameter is opposite to the deviation parameter. The numerical analysis suggests that the inner horizon decreases with increasing $a$ and decreasing $k$. However, the 
outer horizon increases with increasing $a$ and decreasing $k$. 
\section{Thermodynamics}\label{sec03}
Now, we analyze the thermodynamical quantities associated with the   black hole solution (\ref{sol1})  which is characterized by parameters $M$,   $k$ and   $a$.  The black hole mass $M$ can be obtained by setting metric function  $f(r)=0$  at horizon radius $r_+$. This gives   
\begin{equation}
M_+=\frac{(1-a)}{2}\,r_+e^{k/r_+}.
\label{eqM1}
\end{equation}
This expression  exactly coincides with the mass of the  regular black hole for $a=0$, however, this matches with the mass of the Schwarzschild black hole for the vanishing deviation parameter   and CS parameter. 

 The Hawking temperature ($T_+$) evaluated at
the horizon is directly related  to the    surface gravity  $\kappa$  given by  \cite{Singh:2017qur,Chaturvedi:2016fea}
\begin{equation} 
\kappa= \left(-\frac{1}{2}\nabla_{\mu}\xi_{\nu}\nabla^{\mu}\xi^{\nu}\right)^{1/2}=\frac{1}{2}f'(r_+),
\label{temp21}
\end{equation}
and $\xi^{\mu}=\partial/\partial t$ is a Killing vector. 
This form of Hawking temperature  can also be derived from the tunneling method \cite{Maluf:2018lyu}. Corresponding to black hole solution (\ref{sol1}),  the  Hawking temperature is calculated by
\begin{equation}
T_+=\frac{ (1-a)}{4\pi r_+}\left[1-\frac{k}{r_+}\right].
\label{eqT2}
\end{equation}
From the above expression,  it is obvious that, for $k=0$ and $a=0$, the Hawking temperature reduces to $T_+=1/4\pi r_+$ which is exactly identified with the case of Schwarzschild black holes.

Next, we would like to study the entropy of our black hole solution. In general, the explicit form of the entropy $S_+$ can be deduced from the first-law of thermodynamics \cite{ghosh8,Bardeen:1973gs}
\begin{equation}
dM =T_+\,dS.\label{first}
\end{equation}
In fact, this leads to 
\begin{eqnarray}
S_+ =\int\,\frac{1}{T_+}\frac{\partial M }{\partial r_+}dr_+.
\label{eqS1}
\end{eqnarray}
Plugging the values of Mass (\ref{eqM1}) and temperature (\ref{eqT2}) into (\ref{eqS1}), we obtain the entropy of our black hole as
\begin{equation}
S_+ = \pi \left(r_+(k+r_+)e^{k/r_+}-k^2 \text{Ei}\left[\frac{k}{r_+}\right]\right),
\label{entropy1}
\end{equation}
where the last term is characterized by exponential integral.
This entropy does not match with the area-law. However,  in the absence of $k
$, this expression  exactly  matches with  the entropy of Schwarzschild black hole. From the expression (\ref{entropy1}), it is obvious that the area-law is no longer valid for non-singular black holes for large value of $k$.  Now,  we can also  cross-check  the expression of  Hawking temperature evaluated from the 
first-law of thermodynamics (\ref{first}). Here, we have
\begin{equation}
T_+=\frac{\partial M_+}{\partial S_+}=\frac{(1-a)}{4\pi r_+}\Big[1-\frac{k}{r_+}\Big]e^{-k/r_+}.
\label{tempf}
\end{equation}
Here, we see that this expression is not in agreement with one calculated by area-law or tunneling method in (\ref{eqT2}).  
Since expressions of black hole  temperature   (\ref{eqT2}) and  (\ref {tempf}) are estimated  by two different methods.
This means that  the first-law of thermodynamics (\ref{first}) is not
appropriate for deriving the   black hole temperature for regular black holes.  The modified form of first-law for  black holes
is presented in Refs.  \cite{ma14,dvs19}. This modification depends on the general structure of the energy-momentum tensor. In fact, the conventional  first-law modifies with an additional factor if the black hole mass parameter  is included in the energy-momentum tensor.  The modified first-law is, therefore, given as \cite{Maluf:2018lyu, ma14,dvs19}
\begin{equation}
 \mathcal{J}(M,r_+)\,dM=T_+ \,dS,\label{cor}
\end{equation}
where correction factor $ \mathcal{J}(M,r_+)$ is defined in terms of  energy density $T^0_0$ as
\begin{equation}
 \mathcal{J}(M,r_+)=1+4\pi \int_{r_+}^{\infty}r_+^2\frac{\partial T^0_0}{\partial M} dr_+ =e^{-k/r_+}.
\end{equation}
The correction factor in our case is calculated by
\begin{equation}
 \mathcal{J}(M,r_+) = e^{-k/r}.
\end{equation}
The entropy following  the corrected first-law of thermodynamics   (\ref{cor}) is derived as 
\begin{equation}
S_+=\pi r_+^2=\frac{A}{4}.
\label{modent}
\end{equation}
Obviously, this entropy    is in the agreement of the area-law and matches exactly with    the entropy of  black holes.

The thermodynamical stability of  black holes can be understood   by studying the  heat capacity  of the black hole and, therefore, this is calculated by \cite{Chaturvedi:2016fea,ghosh8}
\begin{eqnarray}
C_+=\frac{\partial M_+ }{\partial T_+}=-4\pi r_+^2e^{k/r_+}\frac{(1-k)}{2k-r_+}.
\label{eqC}
\end{eqnarray}
 It is interesting to note that the  heat capacity obtained in the above equation is independent of a CS parameter. This suggests that the stability/instability of the present black hole is not affected by the presence of CS. 
 We also check a second-order phase transition  (Davies point)  in the above-obtained solution, where the temperature of the black hole is  maximum $(dT_+/dr_+=0)$. The heat capacity shows 
 two opposite  behaviors: one of them is { negative}  in the region $r_+<r_C$ (critical radius) which justifies  the thermodynamic { instability} of black hole and, for the region $r_+ > r_C$, this is { positive} which warrants the {  black hole stability}. In fact,   the heat capacity is discontinuous  at $r_+=r_C$ and  divergence occurs, which conforms  the existence of second-order phase transition   \cite{hp,davis77}. If the Davies point   appears then the relationship between  QNMs and phase transition still exists \cite{Jafarzade:2020ova}.

\begin{figure*}[ht]
\begin{tabular}{c c c c}
\includegraphics[width=0.75\linewidth]{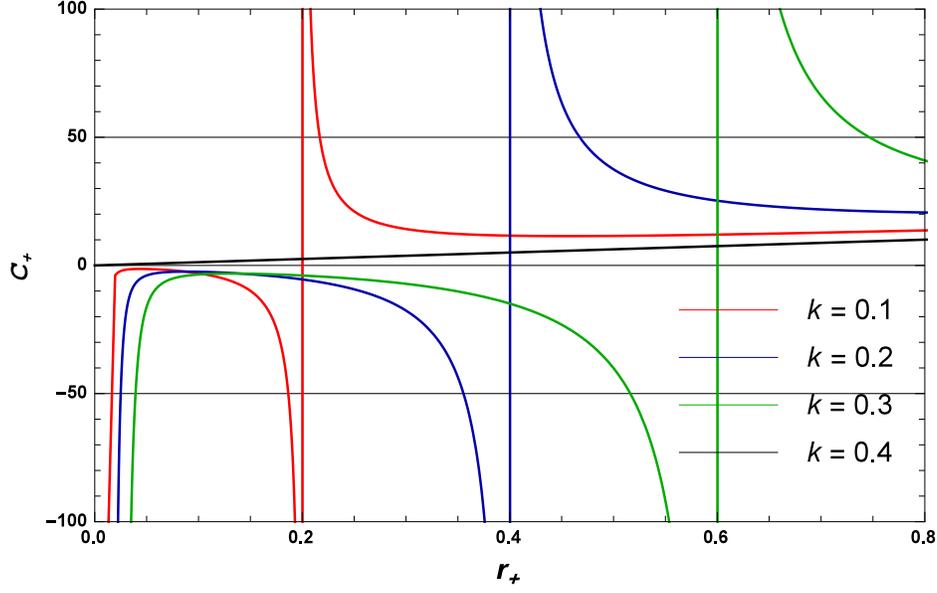}
\end{tabular}
\caption{The plot of specific heat versus horizon radius $r_+$ for different value of deviation parameter $k$.}
\label{rem60}
\end{figure*}
 The { Davies point} occurs where the denominator of the heat capacity (\ref{eqC})  ceases. This gives $r_+=2k$. This confirms that the Davies point occurs where the values of deviation  parameter is half of the horizon radius. 

\section{Null geodesics and photon sphere}\label{sec04}
Null geodesics are  very important to explain
the QNMs of a black hole solution \cite{ch}. 
Here, we follow the photon ring/QNM correspondence \cite{ba,ba1} to  calculate
QNMs of this black hole solution in the eikonal limit.  To compute the geodesics in   spacetime, we first focus on the motion of photons in the black hole solution (\ref{sol1}).  The Lagrangian for a photon motion limited to equatorial plane $(\theta=\pi/2)$ can be expressed as
\begin{equation}
{\cal {L}}=-g_{tt}{\dot t}^2 +g_{rr}{\dot r}^2+g_{\theta\theta}{\dot \theta^2}+g_{\phi\phi} {\dot \phi}^2,
\label{lag1}
\end{equation}
 and the corresponding Hamiltonian is written as
 \begin{equation}
{\cal{H}}=p_t\dot{t}+p_r \dot{r}+p_\phi\dot{\phi}-{\cal {L}} 
 \end{equation}
where over-dot refers to the derivative with respect to an affine parameter. The generalized momenta are given by
\begin{eqnarray}
&&p_t=\frac{\partial {\cal H}}{\partial {\dot t}}\equiv E=\text{constant},\\
&&P_{\phi}=\frac{\partial {\cal H}}{\partial {\dot \phi}}\equiv -J=\text{constant},\\
&&  p_r=\frac{\partial {\cal H}}{\partial {\dot r}}=g_{rr}{\dot r}
\end{eqnarray}
The equation of motion  associated with the photon can be calculated using the Hamiltonian formalism, which can be derived as follow
\begin{equation} 
{\dot t}=\frac{\partial {\cal H}}{\partial p_t}=-\frac{p_t}{g_{tt}},\qquad\qquad {\dot \phi}=\frac{\partial {\cal H}}{\partial p_{\phi}}=\frac{p_{\phi}}{r^2},\quad\text{and}\quad {\dot r}=\frac{\partial {\cal H}}{\partial p_r}=-\frac{p_r}{g_{rr}}.
\end{equation}
The Hamiltonian is independent of  time $(t)$ and azimuthal coordinate $(\phi)$. The null geodesics equation is given by
\begin{equation}
    {\dot r}^2+V_{eff}(r)=0, \qquad \text{with} \qquad V_{eff}=f(r)\left(\frac{J^2}{r^2}-\frac{E^2}{f(r)}\right).
\end{equation}
For  circular null geodesic  which describes the radius of the photon sphere, the effective potential must follow the conditions:
\begin{equation}
V_{eff}=  \frac{\partial V_{eff}}{\partial r}=0.
\label{pot}
\end{equation}
 The equation of the photon radius ($r_p$) is calculated as
\begin{equation}
\frac{2(kM-r_p(3M-(1-a)e^{k/r}r))}{r^2_p(2M-(1-a)e^{k/r}r)} = 0.
\label{pr}
\end{equation}
This  equation can not be solved analytically, so we solve it numerically.   

We can determine the angular radius of the black hole shadow by setting $r$ to $r_p$  (circular orbit of the photon) in Eq. (\ref{pot}). The shadow radius for this black hole  is calculated by
\begin{equation}
r_s=\sqrt{\alpha^2+\beta^2}=\frac{L_p}{E}=\frac{r}{\sqrt{f(r)}}|_{r=r_p}.
\label{sr1}
\end{equation}
Substituting the value of $f(r)$ from Eq. (\ref{sol1}) into Eq. (\ref{sr1}), then we plug the numerical value of $r_p$  in the obtained equation and the numerical results are presented in TABLE \ref{pr1} and TABLE \ref{pr2}. It is noticed  from the TABLES  that the photon radius increases with the deviation parameter and decreases with the CS parameter.

The shadow of the black hole can be  visualized  with the help of celestial coordinates $\alpha$ and $\beta$ \cite{Kumar:2019pjp,Singh:2017vfr,Ahmed:2022qge,Ahmed:2020dzj}. For the obtained  black hole solution the  $\alpha$ and $\beta$  are given by
\begin{eqnarray}
\alpha=\lim_{r\to \infty}\left(\frac{rp^{\phi}}{p^t}\right),\qquad\text{and}\qquad
\beta= \lim_{r\to \infty}\left(\frac{rp^{\theta}}{p^t}\right).
\end{eqnarray}
 
\begin{center}
\begin{table}[ht]
\begin{center}
\begin{tabular}{| l | l| r| l| r| l| r| l |r |r| r  }
\hline
\multicolumn{1}{|c| }{ \it{} } &\multicolumn{1}{c}{ } &\multicolumn{1}{c}{$r_p$  } & \multicolumn{1}{c}{  }& \multicolumn{1}{c|}{} &\multicolumn{1}{c}{ }&\multicolumn{1}{c}{} &\multicolumn{1}{c}{ $r_s$}   & \multicolumn{1}{c|}{}\\
\hline
\multicolumn{1}{ |c|}{ $k$ } &\multicolumn{1}{c|}{ $a=0.1$} &\multicolumn{1}{c|}{$a=0.2$ } & \multicolumn{1}{c|}{ $a=0.3$ }& \multicolumn{1}{c|}{$a=0.4$} &\multicolumn{1}{c|}{ $a=0.1$}&\multicolumn{1}{c|}{$a=0.2$} &\multicolumn{1}{c|}{ $a=0.3$}   & \multicolumn{1}{c|}{$a=0.4$} \\
\hline
\,   $ 0.0$\,&\, 3.333\,  & 3.751\,  & \, 4.285\, &   \, 5.001\, &\, 0.544\, & \, 0.516\, &\, 0.482\, &  \,0.447\, 
 \\
\,   $ 0.1$\,&\, 3.195\,  & 3.612\,  & \, 4.148\, &   \, 4.863\, &\, 0.541\, & \, 0.511\, &\, 0.478\, &  \,0.443\, 
 \\
\,   $ 0.2$\,&\, 3.047\,  & 3.466\,  & \, 4.004\, &   \, 4.720\, &\, 0.534\, & \, 0.505\, &\, 0.474\, &  \,0.440\, 
 \\
\,   $ 0.3$\,&\, 2.886\,  & 3.309\,  & \, 3.851\, &   \, 4.570\, &\, 0.529\, & \, 0.497\, &\, 0.468\, &  \,0.436\, 
 \\
\,   $ 0.4$\,&\, 2.710\,  & 3.139\,  & \, 3.687\, &   \, 4.412\, &\, 0.513\, & \, 0.480\, &\, 0.461\, &  \,0.431\, 
 \\
\,   $ 0.5$\,&\, 2.512\,  & 2.953\,  & \, 3.511\, &   \, 4.244\, &\, 0.497\, & \, 0.477\, &\, 0.453\, &  \,0.425\, 
 \\
\,   $ 0.6$\,&\, 2.283\,  & 2.745\,  & \, 3.318\, &   \, 4.065\, &\, 0.476\, & \, 0.463\, &\, 0.443\, &  \,0.418\, 
 \\
\,   $ 0.7$\,&\, 1.999\,  & 2.505\,  & \, 3.105\, &   \, 3.872\, &\, 0.441\, & \, 0.440\, &\, 0.431\, &  \,0.410\, 
 \\
\hline
\end{tabular}
\end{center}
\caption{The numerical values of photon radius corresponding to the  deviation parameter $(k)$ with fixed   CS parameter $(a)$.}
\label{pr1}
\end{table}
\end{center}
\begin{center}
\begin{table}[ht]
\begin{center}
\begin{tabular}{| l | l| r| l| r| l| r| l |r |r| r  }
\hline
\multicolumn{1}{|c| }{ \it{} } &\multicolumn{1}{c}{ } &\multicolumn{1}{c}{$r_p$  } & \multicolumn{1}{c}{  }& \multicolumn{1}{c|}{} &\multicolumn{1}{c}{ }&\multicolumn{1}{c}{} &\multicolumn{1}{c}{ $r_s$}   & \multicolumn{1}{c|}{}\\
\hline
\multicolumn{1}{ |c|}{ $a$ } &\multicolumn{1}{c|}{ $k=0.1$} &\multicolumn{1}{c|}{$k=0.2$ } & \multicolumn{1}{c|}{ $k=0.3$ }& \multicolumn{1}{c|}{$k=0.4$} &\multicolumn{1}{c|}{ $k=0.1$}&\multicolumn{1}{c|}{$k=0.2$} &\multicolumn{1}{c|}{ $k=0.3$}   & \multicolumn{1}{c|}{$k=0.4$} \\
\hline
\,   $ 0.0$\,&\, 2.8615\,  & 2.711\,  & \, 2.587\, &   \, 2.453\, &\, 0.570\, & \, 0.560\, &\, 0.556\, &  \,0.553\, 
 \\
\,   $ 0.1$\,&\, 3.1954\,  & 2.041\,  & \, 2.926\, &   \, 2.790\, &\, 0.541\, & \, 0.532\, &\, 0.530\, &  \,0.528\, 
 \\
\,   $ 0.2$\,&\, 3.6126\,  & 3.460\,  & \, 3.348\, &   \, 3.220\, &\, 0.511\, & \, 0.505\, &\, 0.502\, &  \,0.501\, 
 \\
\,   $ 0.3$\,&\, 4.1488\,  & 4.004\,  & \, 3.889\, &   \, 3.764\, &\, 0.478\, & \, 0.474\, &\, 0.472\, &  \,0.471\, 
 \\
\,   $ 0.4$\,&\, 4.8636\,  & 4.720\,  & \, 4.619\, &   \, 4.490\, &\, 0.444\, & \, 0.440\, &\, 0.439\, &  \,0.438\, 
 \\
\,   $ 0.5$\,&\, 5.8641\,  & 5.721\,  & \, 5.612\, &   \, 5.490\, &\,0.405\, & \, 0.402\, &\, 0.402\, &  \,0.401\, 
 \\
\,   $ 0.6$\,&\, 7.3647\,  & 7.225\,  & \, 7.117\, &   \, 7.005\, &\, 0.363\, & \, 0.361\, &\, 0.361\, &  \,0.361\, 
 \\
\,   $ 0.7$\,&\, 9.8652\,  & 9.727\,  & \, 9.621\, &   \, 9.513\, &\, 0.315\, & \, 0.313\, &\, 0.313\, &  \,0.313\, 
 \\
\hline
\end{tabular}
\end{center}
\caption{The numerical values of photon radius corresponding to the CS parameter $(a)$ with fixed    deviation parameter $(k)$ .}
\label{pr2}
\end{table}
\end{center}

\begin{figure*}[ht]
\begin{center}
\begin{tabular}{c c c c}
\includegraphics[width=0.4\linewidth]{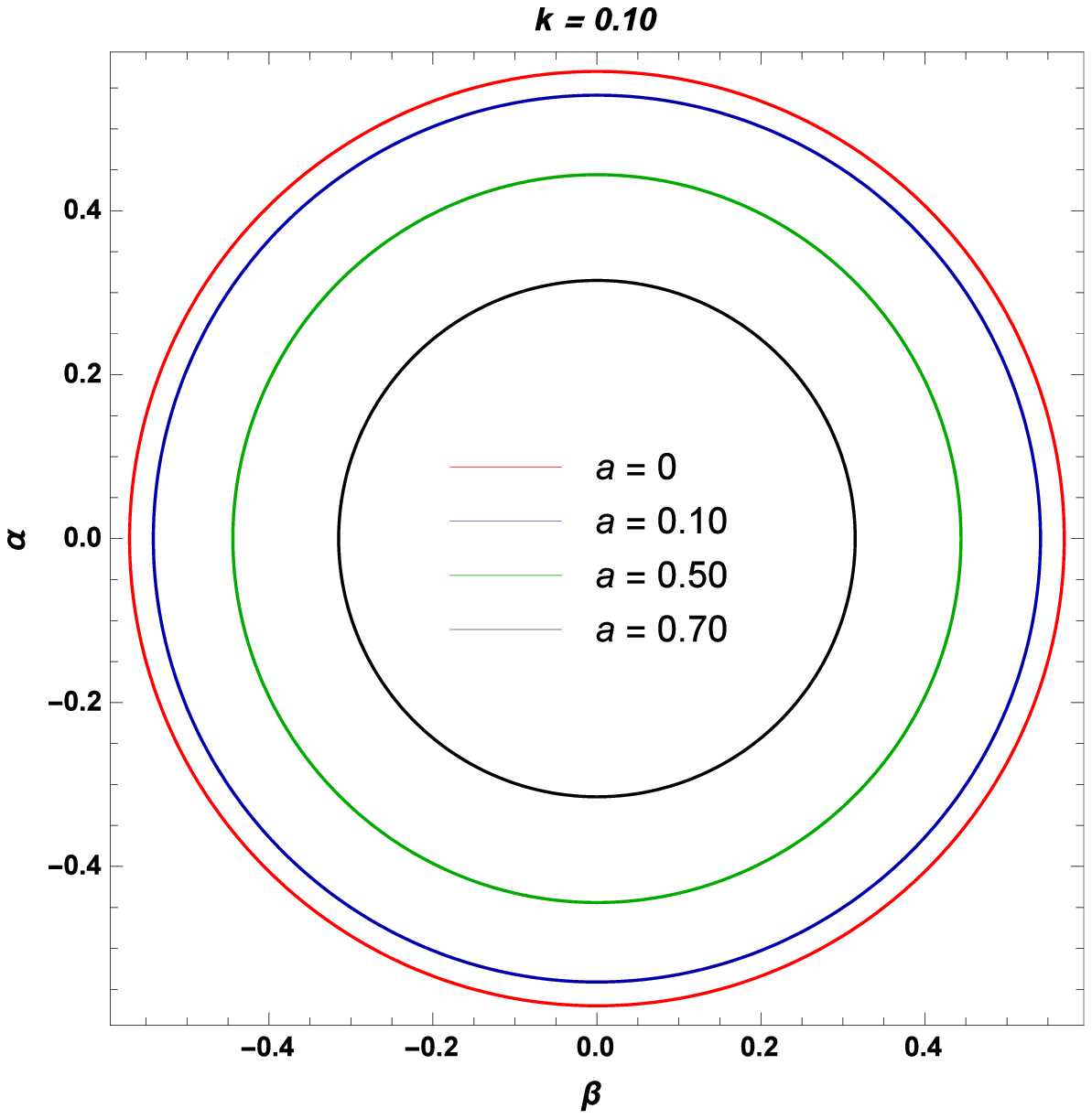}
\includegraphics[width=0.4\linewidth]{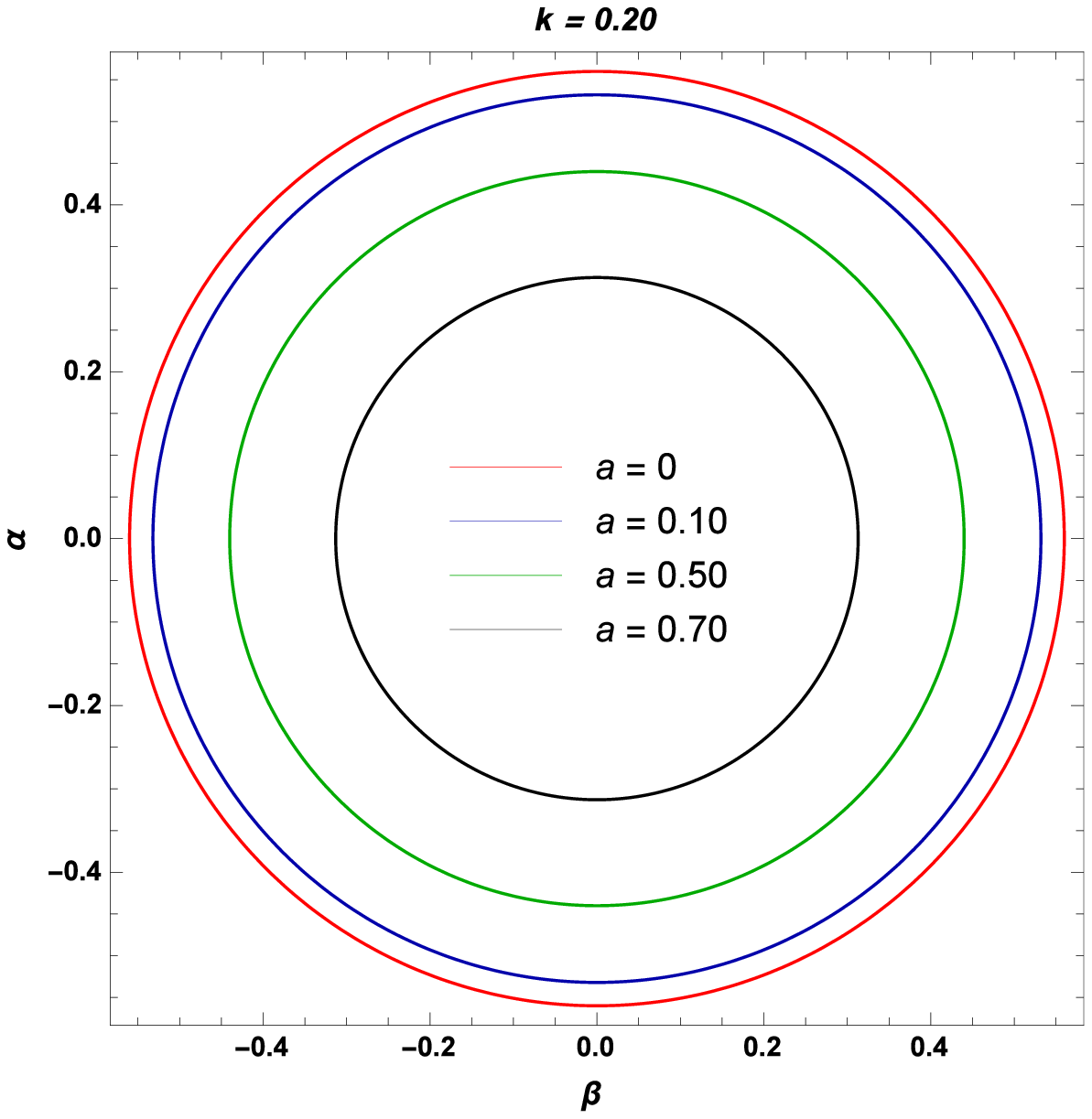}\\
\includegraphics[width=0.4\linewidth]{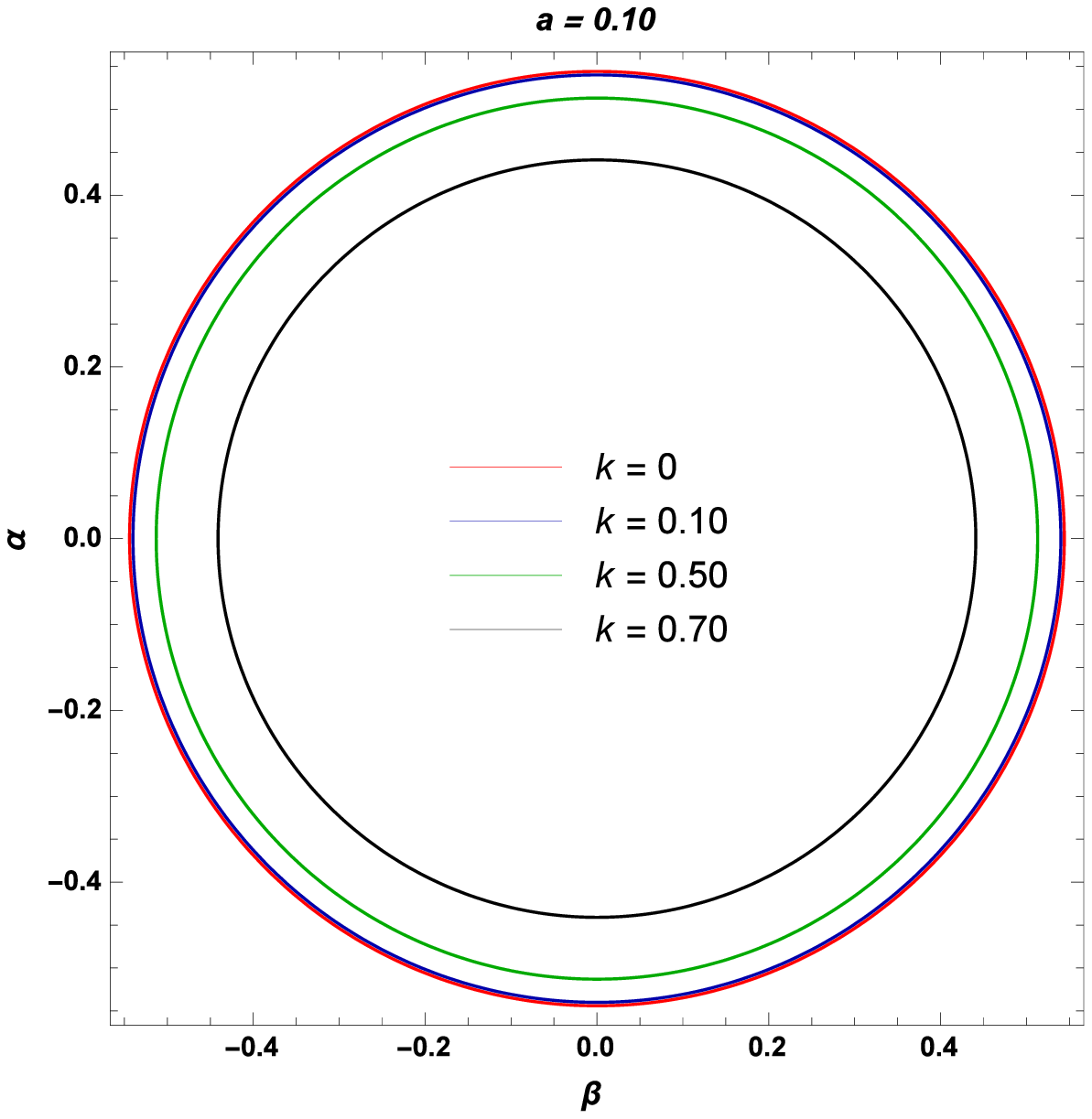}
\includegraphics[width=0.4\linewidth]{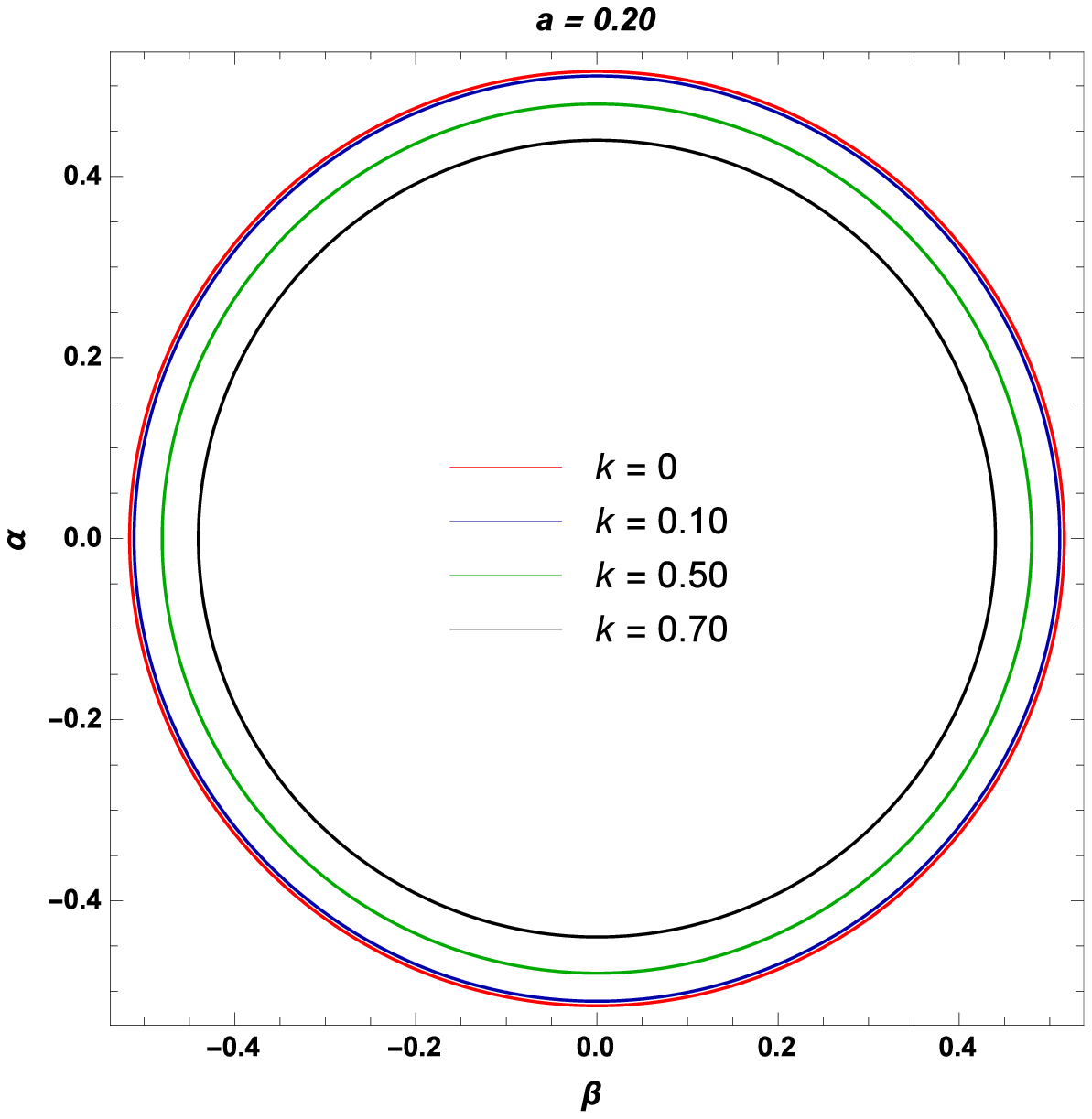}
\end{tabular}
\end{center}
\caption{The plot of black hole shadow  corresponding to the CS  parameter $(a)$ with fixed  deviation parameter $(k)$  (upper panel) and  corresponding to the deviation parameter $(k)$ with fixed CS  parameter $(a)$ (lower panel). }
\label{rem6}
\end{figure*}

The shadow of the obtained solution  for different values of CS parameter $(a)$ and the different values of deviation parameter $(k)$ is depicted in Fig. \ref{rem6}. 
\begin{figure*}[ht]
\begin{tabular}{c c c c}
\includegraphics[width=0.5\linewidth]{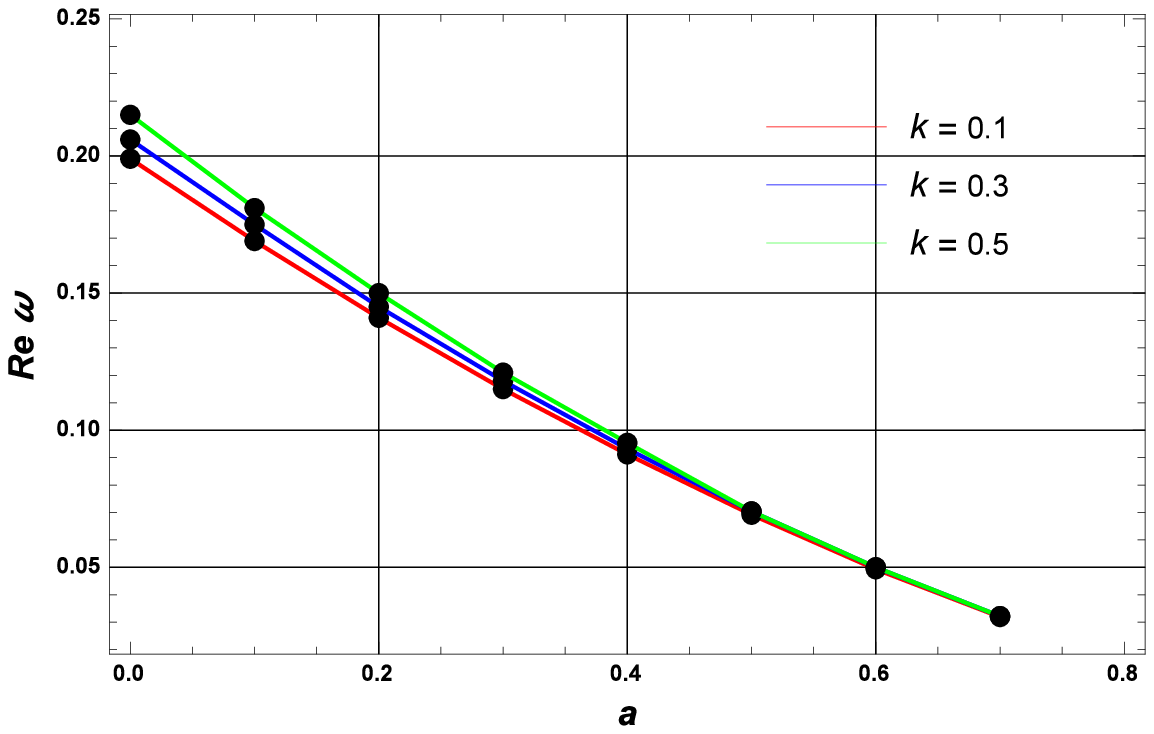}
\includegraphics[width=0.5\linewidth]{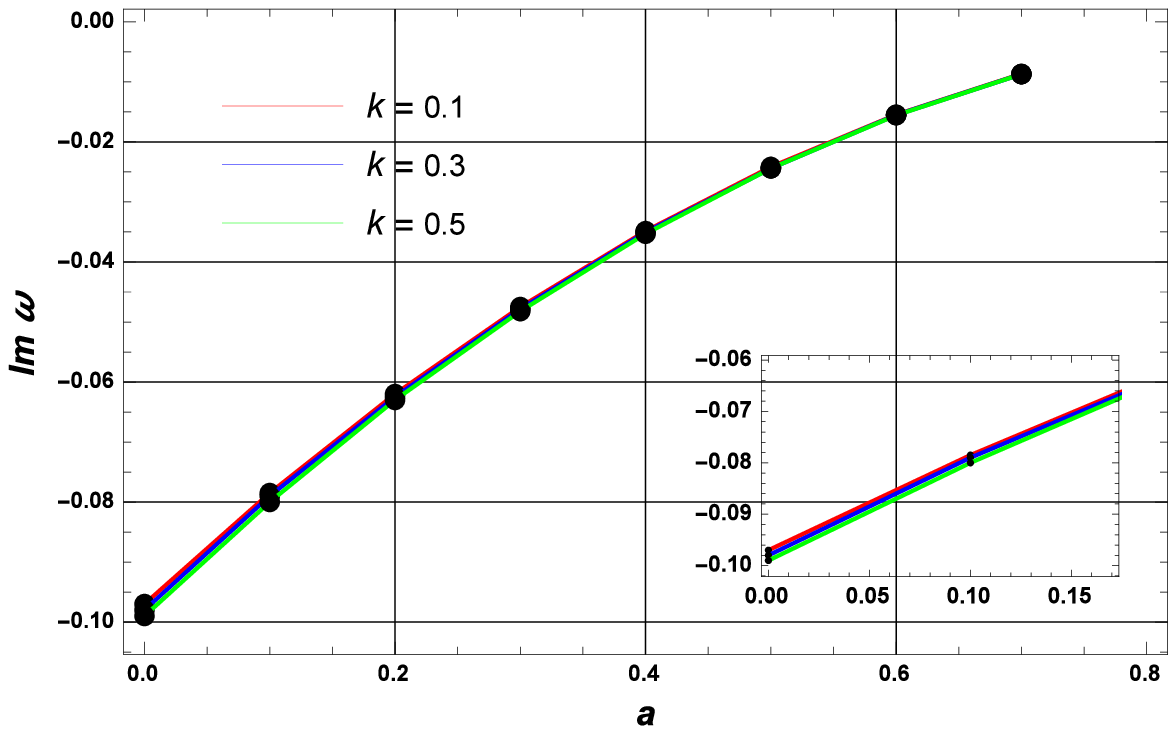}\\
\includegraphics[width=0.5\linewidth]{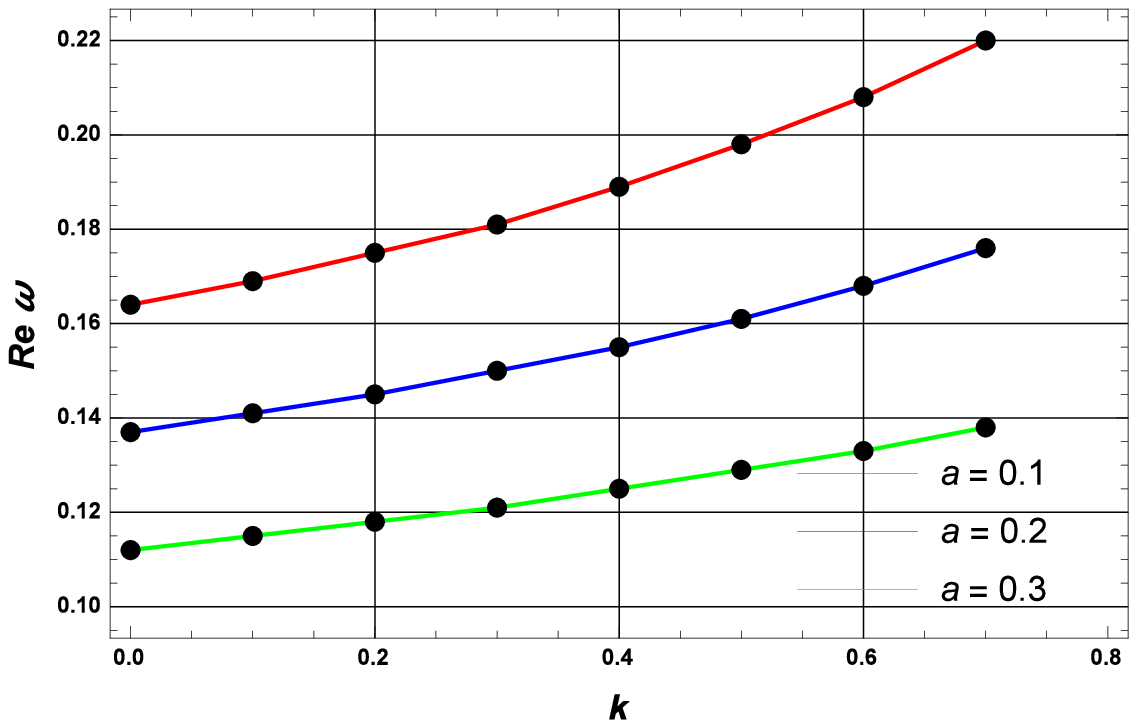}
\includegraphics[width=0.5\linewidth]{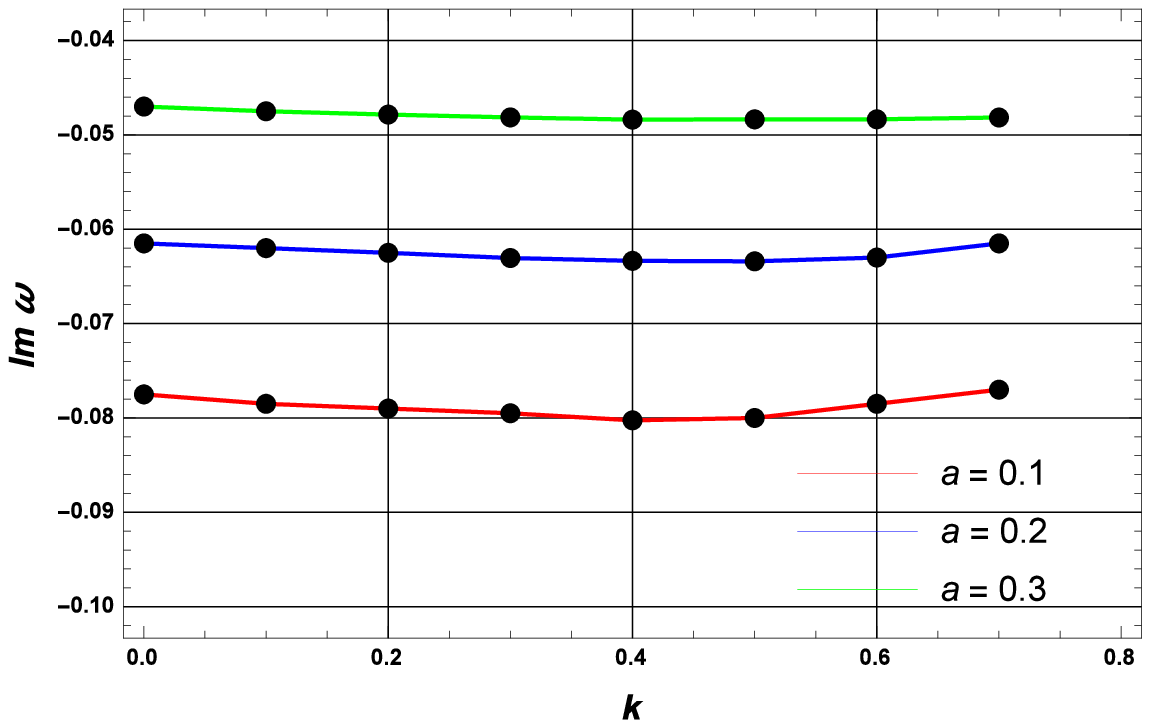}
\end{tabular}
\caption{Upper panel: the real and imaginary part of QNMs with respect to $a$ for different $k$. Lower panel:  the real and imaginary part of QNMs with respect to $k$ for different $a$. }
\label{rem7}
\end{figure*}
From the figure we see that the size of the shadow decreases with the increasing   CS parameter and deviation parameter (as evident from  TABLE \ref{pr1} and TABLE \ref{pr2} also).
As we know the black hole in the limit $r>>k$, the given black hole corresponds to the Reissner-Nordstr\"om black hole 
with CS parameter. So, we can easily comment that  the  shadow  of the resulting black hole with large horizon radius and  small value of $k$ will  only match to the shadow of the Reissner-Nordstr\"om black hole. However, as the value of deviation parameter will increase the shadow radius of the given black hole will get smaller comparative to    the Reissner-Nordstr\"om black hole case. 

\section{Connection beteween shadow radius and QNMs}\label{sec05}

{  The QNMs encode  the details about the stability of the black holes under small perturbation and these are described by the $\omega = \omega_R + i\omega_I$. The signature of the $\omega_I$ characterizes the stability of the black hole. For instance,   $\omega_I>0$ corresponds to stable modes of black holes and the rest corresponds to unstable modes. The real part and imaginary part  of the QNMs in the eikonal limit are associated with the angular velocity and  the
  Lyapunov exponent of (unstable) circular null geodesics, respectively \cite{76,77}. Such a connection also occurs for the QNMs and gravitational lensing \cite{78}. The real part of
QNMs in the eikonal limit is  associated with the radius of the black hole shadow \cite{79, 80}. Such connections  had already been explored to  various black holes \cite{81,82,83}.

\noindent The QNMs frequency $\omega$ can be estimated through  photon sphere as
\begin{equation}
\omega=l\Omega-i\left(n+\frac{1}{2}\right)|\Lambda|,    
\end{equation}
where $n$ is the overtone number and $l$ is the angular quantum number. Here, the angular velocity $(\Omega)$ and the Lyapunov exponent $(\Lambda)$ of the photon sphere are given, respectively, by
\begin{eqnarray}
 &&\Omega=\frac{\sqrt{f(r_p)}}{r_p}=\frac{1}{L_p},\\
&& \Lambda=\frac{\sqrt{f(r_p)(2f{r_p}-r^2_pf''(r_p))}}{\sqrt{2} r_p}.
\end{eqnarray}
{ The QNMs are characterized by  complex numbers. The imaginary part of the QNMs tells about the stability of the black hole such that when $Im\ \omega >0$  the system is stable and when $Im\ \omega <0$ the system is unstable. The real   and imaginary parts  of the QNMs for the given black hole  (\ref{sol1}) for different values of deviation parameter $(k)$ and CS parameter $(a)$  are depicted  in the figure \ref{rem7}. From  the figure, we make the following observations: \begin{enumerate}
 \item The upper panel of  figure \ref{rem7} tells that the real (imaginary) part of the QNMs frequency decreases (increases) with increasing the CS parameter.  This indicates that the scalar field perturbation in the presence of deviation parameter oscillates faster and decays slower as compared to Reisner-Nordstr\"om  black hole \cite{Leaver:1990zz}. 
\item The lower panel of  figure \ref{rem7} suggests that both the real  and imaginary parts of the QNMs frequency   increase with the   deviation parameter. This means that the scalar field perturbation in the presence of CS parameter oscillates and decays faster as compared to Reisner-Nordstr\"om  black hole \cite{Leaver:1990zz}.
\end{enumerate}
Here, it is worthwhile to mention that  the effects of $k$ and $a$ are opposite to each other. From the right panel of figure \ref{rem7}, we notice that the $Im \,\omega$ is negative for both $k$ and $a$. Therefore, the obtained black hole solution is stable.
}
\section{Conclusions}\label{sec5}
The nonlinear electrodynamics as a source of gravity has the possibility to create   
regular black
holes   and star (or soliton)  like configurations of interest. 
In this work, we have focused on a   gravity model coupled with the
nonlinear electrodynamics (which resembles  the electrodynamics in the weak 
field
limit) in the presence of a cloud of strings as the source. The dynamics of such the
gravity model is described by a suitable  action and corresponding  equation 
of motions. 
 
Furthermore, we have discussed the thermodynamics of the black hole solution. 
The effects of the  deviation
parameter   and CS parameter on the temperature  are emphasized. The temperature is a decreasing function of the  deviation
parameter   and CS parameter.
Hawking temperature of the black hole is 
Interestingly, here we have found that  instead of the standard first-law of thermodynamics this black hole   follows a modified first-law of  thermodynamics. From the specific heat, we have found that there is no effect of  surrounding  CS on the stability of black hole. However, the second-order phase transition  occurs at certain point (Davies point),  where deviation parameter is half of the value of horizon radius. 
The existence of Davies point confirms the relationship between QNMs and phase transition.     QNMs  for this particular black hole solution in the eikonal 
limit   are calculated under the consideration of photon ring/QNM correspondence.  The photon radius  is obtained numerically. 
There are many interesting opportunities that can be studied further in future
works. For example, it will be interesting to explore the  quantum effects on the thermodynamics and stability of the regular  black hole solution.
 
 We also study the relation between the shadow radius and QNMs of the solution (\ref{sol1}). From the figures  \ref{rem6} and \ref{rem7}, we found that the real  and imaginary  part  of the QNMs decreases   and increases 
with   the increasing the CS parameter, respectively. However, and the   shadow radius  decreases with the increasing the CS parameter. However, the real and imaginary part  of QNMs   increases  with increasing deviation parameter  and shadow radius    decreases with increasing deviation parameter. 
The shadow behavior  and photon sphere in the presence of black hole parameter $k$ and $a$ are also studied. The numerical results (cf. TABLE \ref{pr1}) confirmed that the photon sphere $(r_p)$ and shadow radius $(r_s)$ are decreasing with the deviation parameter. However,  with the CS parameter (cf. TABLE \ref{pr2}), the photon radius ($r_p$) is increasing and shadow radius ($r_s$) is decreasing. 
 
\section*{Data Availability Statement} 
Data sharing not applicable to this article as no datasets were generated or analysed during the current study.


\begin{thebibliography}{99}
\bibitem{st} H. Stephani, D. Kramer, M. MacCallum  and C. Hoenselaers, Exact 
Solutions of Einstein’s Field Equations
(Cambridge University Press, second edition, Cambridge
2002).
\bibitem{bar}J. M. Bardeen, in Proceedings of GR5 (Tbilisi, URSS,
1968).
\bibitem{reg} I. G. Dymnikova, Gen. Relativ. Gravit. 24, 235 (1992); Int. J. Mod. Phys. D 05, 529 (1996); Int. J. Mod. Phys. D 12, 1015 (2003).
\bibitem{reg1}  K. A. Bronnikov, Phys. Rev. D 63, 044005 (2001).
\bibitem{reg2}  S. A. Hayward, Phys. Rev. Lett. 96, 031103 (2006).
\bibitem{gla} T. Tangphati, A. Pradhan, A. Banerjee and G. Panotopoulos, Phys. Dark Univ. 33 (2021) 100877.
\bibitem{gla1} J. M. Z. Pretel, A. Banerjee and A. Pradhan,
Eur. Phys. J. C 82 (2022) 180.
\bibitem{gla2} T. Tangphati, A. Pradhan, A. Errehymy and A. Banerjee, Phys. Lett. B 819 (2021) 136423.
\bibitem{gla3}
R.~P.~Singh, B.~K.~Singh, B. R.~K.~Gupta and S.~Sachan,
 Can. J. Phys. 100, 39 (2022).
\bibitem{na}J. C. S. Neves, Int. J. Mod. Phys. A 32, 1750112 (2017).
\bibitem{born} M. Born and L. Infeld, Proc. Roy. Soc. Lond. A 144, 425 (1934).
\bibitem{b1} D. L. Wiltshire,  Phys. Rev. D38, 2445 (1988).
 \bibitem{b2} T. Tamaki and  T. Torii, Phys. Rev. D62, 061501 (2000).
 \bibitem{b3}  N.  Breton,  Phys. Rev. D 67, 124004 (2003).
 \bibitem{b4} S.  Fernando and  D. Krug, Gen. Rel. Grav. 35, 129 (2003). 
\bibitem{b5} R. G.  Cai, D. W.  Pang and   A. Wang, Phys. Rev. D70, 124034 (2004).
\bibitem{b6} A. A.  Tseytlin,  Nucl. Phys. B 276, 391 (1986).
 \bibitem{b7} L, De Fosse,  P. Koerber and  A. Sevrin, Nucl. Phys. B 603, 413 (2001).
 \bibitem{lat} P. S. Letelier,   Phys. Rev. D 20, 1294 (1979).
\bibitem{lat1}P. S. Letelier, Il Nuovo Cim. B  63, 519 (1981). 
\bibitem{lat2} P. S. Letelier, Phys. Rev. D 28, 2414 (1983).
\bibitem{cs1} 
 A. Ganguly, S. G. Ghosh and S. D. Maharaj,   Phys. Rev. D 90,  064037 (2014).
\bibitem{cs2}  K. A. Bronnikov, S. W. Kim and M. V. Skvortsova,   Class. Quant. Grav. 33, 195006 (2016).
\bibitem{cs3}  D. Barbosa and V. B. Bezerra,   Gen. Rel. Grav. 48, 149 (2016). 
\bibitem{cs4}  E. Herscovich and M. G. Richarte,  Phys. Lett. B 689, 192 (2010).
\bibitem{cs5}  S. G. Ghosh and S. D. Maharaj, Phys. Rev. D 89,   084027 (2014).
 \bibitem{cs6} S. H. Mazharimousavi and M. Halilsoy,   Eur. Phys. J. C 76,   95 (2016).
\bibitem{synge} J. L. Synge, \emph{Relativity: The General Theory}, (North Holland, Amsterdam, 1966), p. 175.
\bibitem{mb}
M. Barriola and A. Vilenkin, Phys. Rev. Lett. { 63}, 341 (1989).
 \bibitem{chan}S. Chandrasekhar and S. Detweller, Proc. R. Soc. Lond. A 344, 441 (1975).
 \bibitem{chan01}J. Jing and Q. Pan, Phys. Lett. B 660, 13 (2008).
 \bibitem{76}
V. Cardoso, A. S. Miranda, E. Berti, H. Witek and V.T. Zanchin,  Phys. Rev. D { 79}, 064016 (2009).
 \bibitem{Jafarzade:2020ova}
K.~Jafarzade, M.~Kord Zangeneh and F.~S.~N.~Lobo,
JCAP  {04} (2021)  008.
 \bibitem{sudhak}S. Mandal, S. Upadhyay, Y. Myrzakulov and G. Yergaliyeva, arXiv:2207.10085.
 
\bibitem {sus} S. G. Ghosh, U. Papnoi, and S. D. Maharaj, Phys. Rev. D { 90}, 044068 (2014).
 
\bibitem{Ghosh:2018bxg}
S.~G.~Ghosh, D.~V.~Singh and S.~D.~Maharaj,
Phys. Rev. D  {97} (2018)  104050.
\bibitem{ak19}
A. Kumar, D. V. singh and S.G. Ghosh, Eur. Phys. Journal C { 79} 275 (2019).
 
\bibitem{sabir}
 Md. Sabir Ali and S. G. Ghosh, Phys. Rev. D { 98}, 084025 (2018).
 

\bibitem{Singh:2017qur}
  D.~V.~Singh and N.~K.~Singh,
  Annals Phys.\  { 383}, 600 (2017).
 
\bibitem{Chaturvedi:2016fea} 
  P.~Chaturvedi, N.~K.~Singh and D.~V.~Singh,
  Int.  J. Mod. Phys.  D { 26}, 1750082 (2017).
  \bibitem{Maluf:2018lyu} 
  R.~V.~Maluf and J.~C.~S.~Neves,
  Phys.  Rev. D { 97}, 104015 (2018).
 
\bibitem{ghosh8}
S. G. Ghosh and D. W. Deshkar  Phys. Rev. D { 77}, 04750.
\bibitem{Bardeen:1973gs}
  J.~M.~Bardeen, B.~Carter and S.~W.~Hawking,
  Commun.  Math. Phys.  { 31} 161 (1973).

\bibitem{ma14}
M.-Sen Ma and R. Zhao, Class. Quantun Grav. { {31}}  245014 (2014).

 \bibitem{dvs19}
D. V. Singh, S. G. Ghosh and S. D. Maharaj, Annals Phys.  {  412}, 168025 (2020).

\bibitem{hp}S. Hawking and D. Page, Commun. Math. Phys. {  87}, 577 (1983).
\bibitem{davis77}
P. Davis, Proc. R. Soc.A {\bf 353}, 499 (1977).
\bibitem{ch} S. Chandrasekhar, The Mathematical Theory of Black
Holes, (Oxford University Press, New York, 1983). 
\bibitem{ba}V. Cardoso, A. S. Miranda, E. Berti, H. Witek, V. T. Zanchin, Phys. Rev. D 79 (2009) 064016.   
\bibitem{ba1} N. Breton, L. A. Lopez,   Phys. Rev. D 94 (10) (2016) 104008.
\bibitem{Kumar:2019pjp}
R.~Kumar, S.~G.~Ghosh and A.~Wang,
Phys. Rev. D  {100} (2019)   124024.
\bibitem{Singh:2017vfr}
B.~P.~Singh and S.~G.~Ghosh,
Annals Phys.  {395} (2018) 127.
\bibitem{Ahmed:2022qge}
F.~Ahmed, D.~V.~Singh and S.~G.~Ghosh, Gen. Rel. Grav.  {54} (2022)  21.
\bibitem{Ahmed:2020dzj}
F.~Ahmed, D.~V.~Singh and S.~G.~Ghosh, 
arXiv:2008.10241 [gr-qc].

\bibitem{77}
 R. A. Konoplya, Z. Stuchlik
Phys. Lett. B {\bf 771}, 597 (2017).
\bibitem{78}
I. Z. Stefanov,  Phys. Rev. Lett. { 104}, 251103 (2010).
\bibitem{79}
 K. Jusufi,  Phys. Rev. D { 101}, 084055 (2020).
\bibitem{80}
 K. Jusufi,
Phys. Rev. D {101}  124063 (2020).
\bibitem{81}
 Y. Guo and Y. G. Miao,
Phys. Rev. D {102}  084057 (2020).
\bibitem{82}
 C. Lan, Y. G. Miao and H. Yang, Nucl. Phys. B {971} (2021) 115539.
\bibitem{83}
 S. W. Wei and Y. X. Liu,
Chin .Phys. C {\bf 44}, 115103 (2020).

\bibitem{Leaver:1990zz}
E.~W.~Leaver,
Phys. Rev. D \textbf{41} (1990)  2986. 
\end{thebibliography}
\end{document}